\documentclass[a4paper,11pt]{article}
\usepackage{jheppub} % for details on the use of the package, please see the JINST-author-manual
\usepackage{lineno}
\usepackage{ifthen}

\newcommand{\alg}[1]{\mathfrak{#1}}
\newcommand{\grp}[1]{\operatorname{#1}}

\newcommand{\GL}[2][]{%
   \ifthenelse{ \equal{#1}{} }
      {\grp{GL}({#2})}
      {\grp{GL}^{#1}({#2})}
}
\newcommand{\U}[2][]{%
   \ifthenelse{ \equal{#1}{} }
      {\grp{U}({#2})}
      {\grp{U}^{#1}({#2})}
}
\newcommand{\uu}[2][]{%
   \ifthenelse{ \equal{#1}{} }
      {\alg{u}({#2})}
      {\alg{u}^{#1}({#2})}
}
\newcommand{\SL}[2][]{%
   \ifthenelse{ \equal{#1}{} }
      {\grp{SL}({#2})}
      {\grp{SL}^{#1}({#2})}
}
\newcommand{\SO}[2][]{%
   \ifthenelse{ \equal{#1}{} }
      {\grp{SO}({#2})}
      {\grp{SO}^{#1}({#2})}
}
\newcommand{\SU}[2][]{%
   \ifthenelse{ \equal{#1}{} }
      {\grp{SU}({#2})}
      {\grp{SU}^{#1}({#2})}
}
\newcommand{\Spin}[2][]{%
   \ifthenelse{ \equal{#1}{} }
      {\grp{Spin}({#2})}
      {\grp{Spin}^{#1}({#2})}
}
\newcommand{\Pin}[2][]{%
   \ifthenelse{ \equal{#1}{} }
      {\grp{Pin}({#2})}
      {\grp{Pin}^{#1}({#2})}
}
\newcommand{\Orth}[2][]{%
   \ifthenelse{ \equal{#1}{} }
      {\grp{O}({#2})}
      {\grp{O}^{#1}({#2})}
}
\newcommand{\PO}[2][]{%
   \ifthenelse{ \equal{#1}{} }
      {\grp{PO}({#2})}
      {\grp{PO}^{#1}({#2})}
}
\newcommand{\PSO}[2][]{%
   \ifthenelse{ \equal{#1}{} }
      {\grp{PSO}({#2})}
      {\grp{PSO}^{#1}({#2})}
}
\newcommand{\SP}[2][]{%
   \ifthenelse{ \equal{#1}{} }
      {\grp{SP}({#2})}
      {\grp{SP}^{#1}({#2})}
}
\newcommand{\Sp}[2][]{%
   \ifthenelse{ \equal{#1}{} }
      {\grp{Sp}({#2})}
      {\grp{Sp}^{#1}({#2})}
}
\newcommand{\SK}[2][]{%
   \ifthenelse{ \equal{#1}{} }
      {\grp{SO}^*({#2})}
      {\grp{SO}^*^{#1}({#2})}
}
\newcommand{\cl}[2][]{%
   \ifthenelse{ \equal{#1}{} }
      {\grp{C\ell}({#2})}
      {\grp{C\ell}^{#1}({#2})}
}
\newcommand{\sk}[2][]{%
   \ifthenelse{ \equal{#1}{} }
      {\alg{so}^*({#2})}
      {\alg{so}^*^{#1}({#2})}
}
\newcommand{\symp}[2][]{%
   \ifthenelse{ \equal{#1}{} }
      {\alg{sp}({#2})}
      {\alg{sp}^{#1}({#2})}
}
\newcommand{\spl}[2][]{%
   \ifthenelse{ \equal{#1}{} }
      {\alg{sl}({#2})}
      {\alg{sl}^{#1}({#2})}
}
\newcommand{\so}[2][]{%
   \ifthenelse{ \equal{#1}{} }
      {\alg{so}({#2})}
      {\alg{so}^{#1}({#2})}
}
\newcommand{\su}[2][]{%
   \ifthenelse{ \equal{#1}{} }
      {\alg{su}({#2})}
      {\alg{su}^{#1}({#2})}
}
\newcommand{\spin}[2][]{%
   \ifthenelse{ \equal{#1}{} }
      {\alg{spin}({#2})}
      {\alg{spin}^{#1}({#2})}
}

\arxivnumber{2606.02477} % if you have one

\title{\boldmath Parity Oddness of $\operatorname{Spin}(1,4)$ Dirac Mass Terms}

\author{Craig M$^{\mathrm{c}}$Rae}
\affiliation{University of Manitoba,\\
66 Chancellors Circle Winnipeg, Canada}

\emailAdd{mcraec3@myumanitoba.ca}

\abstract{In the representation theory of Lorentzian orthogonal groups there are well known arguments as to why the parity operator $\mathcal{P}$ and the time reversal operator $\mathcal{T}$ should be realized as linear and anti-linear operators respectively (Wigner 1932). In this paper it is shown that the only operators satisfying the requisite properties for the spinor representation of the de Sitter group $\operatorname{SO}(1,4)$ lead to fermion self-couplings which are necessarily parity odd, ruling out standard Dirac mass terms for theories with fermions obeying $\operatorname{Spin}(1,4)$ symmetry.}

\begin{document}
\maketitle
\flushbottom

\section{Introduction}
\label{sec:intro}
Standard techniques for dealing with fermions in curved spacetime have no issue placing massive fermions upon a de Sitter background. This is because the fermions locally obey the usual $\Spin{1,3}$ Lorentzian symmetry for which the mass term is given in the Lagrangian by the term $m\overline{\psi}\psi$, and this term is a scalar under the entire Lorentz group $\Orth{1,3}$. The story is different if we demand the fermions instead obey a $\Spin{1,4}$ symmetry, often called the de Sitter group. This was first done by Dirac \cite{DiracdeSit}, who demanded the spinors obey the symmetry as a global spacetime symmetry, and whose work is superseded by approaches to quantum field theory in curved spacetime. There is however another interesting case \cite{dSTangentGrav}, where one has local de Sitter symmetry on a four dimensional Lorentzian manifold, instead of local usual Lorentz symmetry. It is in these latter cases, where the spinors respect the full $\Spin{1,4}$ symmetry group, that a standard Dirac mass term can be seen to be parity odd.

Recall that for Lorentzian orthogonal groups the entire orthogonal group of $\Orth{1,n}$ is given by the semi-direct product: $\Orth{1,n} = \SO[+]{1,n} \rtimes \grp{K}_4$, where $\SO[+]{1,n}$ is the identity component of the group, and $\grp{K}_4 $ is the Klein four group of spacetime reflections: $\{1, \mathcal{P}, \mathcal{T}, \mathcal{PT}\}$. The representation theory of these groups is well understood, however to understand the action of spacetime reflections upon spinors, we must understand representations of double covers of the entire orthogonal group $\Orth{1,n}$. It is well known to mathematicians that one may construct these double covers by building what we call \textit{pin} groups, which fall out of the Clifford algebra approach to building spin groups (i.e. constructing gamma matrices).\footnote{For a great introduction to the study of Clifford algebras and their relationship to spinors and orthogonal groups see Vaz \& Rocha \cite{CliffAlg}.} Pin groups can also be seen as a semi-direct product between the identity component of the relevant spin group and a discrete reflection group (a double cover of $\grp{K}_4$); as such once the relevant spin group is understood, there are only three reflections whose behaviours need to be specified in order to build the relevant pin group: parity ($\mathcal{P}$), time reversal ($\mathcal{T}$), and their product ($\mathcal{PT}$). With all this in mind, it is worth noting finally that the standard Clifford algebra method of constructing double covers of reflection groups is physically insufficient due to the resulting time reversal operators being necessarily linear and not anti-linear. `Representations' which map group elements not only to linear maps, but to anti-linear maps as well, have been dubbed \textit{co-representations} by Wigner \cite{Wigner1932,WIGNERBOOK}. Below a co-representation of $\Pin{4,1}$\footnote{In general $\Pin{p,q} \neq \Pin{q,p}$. Formally one can show $\Pin{4,1}$ is the construction done here. Later in footnote \ref{laternote} it is discussed why this does not matter, as changing signature does not change the results.} is constructed and it is shown that the Dirac mass term $m \overline{\psi}\psi$, with $\psi$ being a generic $\Spin{1,4}$ spinor, must be parity odd.

\section{Building a \texorpdfstring{Co-Representation of $\grp{Pin}(4,1)$}{Co-Representation of Pin(4,1)}}

The de Sitter group $\Orth{1,4}$ is the isometry group of de Sitter spacetime \cite{MTW}, or equivalently the non-translational symmetries of a five dimensional Minkowski spacetime. In \cite{dSTangentGrav} it is merely an extended symmetry group at the tangent spaces of a Lorentzian manifold. Regardless of the system at hand, if $\eta_{AB}$ are the components of the $5\times5$ Minkowski metric with a chosen signature, we may construct our spin and pin groups starting with gamma matrices satisfying the Clifford algebra relation:
\begin{equation}
    \gamma_A\gamma_B + \gamma_B\gamma_A = 2\eta_{AB}.
\end{equation}
The elements of $\alg{spin}(1,4)$ are then spanned by the commutators of $\gamma_A$:
\begin{equation}
    s_{AB} = \frac{1}{4}\left[\gamma_A,\gamma_B\right] ,\quad A, B  \in\{ 0,1,2,3,5 \}.
\end{equation}
The first $6$ generators are the familiar Lorentz group boosts and rotations. The four additional generators give $s_{05} \propto \gamma_0\gamma_5$, which squares to $+1$, and so generates a non-compact `time translation', while $s_{i5}\propto\gamma_i\gamma_5$ are compact and generate, in the global picture, `spacial translation' around the closed de Sitter universe. These interpretations are helpful in thinking about the symmetry, but not required. With $L$ for rotations, $K$ for boosts, and $T$ for `translations', one basis for the generators of $\spin{1,4}$, written in a compact quaternionic form is:
\begin{equation}
\begin{split}
    \begin{Bmatrix}
        L_x \\
        L_y \\ 
        L_z 
    \end{Bmatrix} &= \frac{1}{2}     \begin{Bmatrix}
        i \\
        j \\ 
        k 
    \end{Bmatrix} \begin{pmatrix}
        1 & 0 \\ 
        0 & 1
    \end{pmatrix} ,\quad \begin{Bmatrix}
        K_x \\
        K_y \\ 
        K_z 
    \end{Bmatrix} = \frac{1}{2}     \begin{Bmatrix}
        i \\
        j \\ 
        k 
    \end{Bmatrix} \begin{pmatrix}
        0 & 1 \\ 
        -1 & 0
    \end{pmatrix}, \\
    \begin{Bmatrix}
        T_x \\
        T_y \\ 
        T_z 
    \end{Bmatrix} &= \frac{1}{2}\begin{Bmatrix}
        i \\
        j \\ 
        k
    \end{Bmatrix} \begin{pmatrix}
        1 & 0 \\
        0 & -1
    \end{pmatrix}, \qquad 
    \>\>T_t = \frac{1}{2}\begin{pmatrix}
        0 & 1 \\
        1 & 0
    \end{pmatrix}.
\end{split}
\end{equation}
These can be mapped to the complex setting by the element wise mapping $\{i, j, k\} \mapsto \{i\sigma_x,-i\sigma_y,i\sigma_z\}$. Before discussing reflections we will ask first about the Dirac adjoint for de Sitter symmetry. For all elements of the Lie algebra $\lambda$, the adjoint matrix $h$ should satisfy
\begin{equation}
    \lambda ^\dagger h = -h\lambda.
\end{equation}
One can confirm the only matrix satisfying this is proportional to $\gamma_0$, so the Dirac adjoint is no different for our $\Spin{1,4}$ spinors compared to the standard $\Spin{1,3}$. 

With the spin group understood we merely need to find the discrete operators $\mathcal{P}$, $\mathcal{T}$, and $\mathcal{PT}$. Searching for elements of the Clifford algebra which behave as expected\footnote{`As expected' here means the generators of the transformations have parity and time reversal properties which are self-consistent with the commutator, as well as the group action.} we have that parity $\mathcal{P}$ should commute with rotations, reverse boosts, reverse spacial translations, and commute with time translation. From these constraints only $\gamma_0\gamma_5$ could serve as $\mathcal{P}$. Likewise time reversal $\mathcal{T}$ should commute with rotations, reverse boosts, reverse time translation, and commute with spacial translation. Only $\gamma_0$ satisfies the stated requirements here, however we expect the correct time reversal operator $\mathcal{T}$ in the correct co-representation of the full Pin group to be anti-linear, which none of our gamma matrices are. Thus in order to acquire the requisite anti-linearity we will extend the group generators by the quaternionic structure $J$, given in this basis as $J = \gamma_3 \gamma_1*$, where $*$ is complex conjugation acting to the right. $J$ necessarily commutes with every element of the spin group, and $J^2 = -1$. A such $\gamma^0 J$ has all the correct behaviour to be $\mathcal{T}$: it commutes and anti-commutes the correct spin generators, is anti-linear, and squares to $-1$. This gives a discrete spacetime reflection group generated by the following elements:
\begin{equation}
    \mathcal{P} = \gamma^0 \gamma^5, \quad \mathcal{T} = \gamma^0 J, \quad \mathcal{PT} = \gamma^5 J.
\end{equation}
We have arrived at a satisfactory co-representation of $\Pin{4,1}$, where each element of $\Orth{1,4}$ has (up to a sign) a unique action upon our spinors. Let us see how our possible Lagrangian terms behave when committed to this symmetry.
\section{Parity and Time Reversal Upon \texorpdfstring{$\Spin{1,4}$}{Spin(1,4)} Spinor Bilinears}
It is clear from the spin representations that an outer product of two $\Spin{1,4}$ spinors $\psi$, $\varphi$, decomposes as
\begin{equation}
    \psi\overline{\varphi} = a \mathbb{I}_4 + V^A \gamma_A + G^{AB} s_{AB} = \mathbf{1}+\mathbf{5}+\mathbf{10},
\end{equation} a $1$ dimensional scalar representation, a $5$ dimensional vector representation, and a $10$ dimensional adjoint (anti-symmetric tensor) representation. Investigating the parity and time reversal properties of these products we have:
\begin{equation}
    \mathcal{P}\left[\psi \overline{\varphi}\right] = \mathcal{P}\psi \overline{\mathcal{P}\varphi} = \mathcal{P}\psi \varphi^\dagger\mathcal{P}^\dagger \gamma^0 = \mathcal{P} \left(\psi \overline{\varphi} \right)\overline{\mathcal{P}},
\end{equation}
where $\overline{\mathcal{P}} = \gamma^0 \mathcal{P}^\dagger \gamma^0$, and similarly for $\mathcal{T}$. Carrying out the algebra and keeping track of sign changes of components, the decomposition is seen to become
\begin{equation}
\begin{split}
    \mathcal{P} \left(\psi \overline{\varphi}\right) \overline{\mathcal{P}} &= \left(\gamma_0 \gamma_5\right)\left(a \mathbb{I}_4 + V^A \gamma_A + G^{AB} s_{AB} \right) \gamma_0 \left(\gamma_0 \gamma_5\right) \gamma_0, \\
    &= a^\prime \mathbb{I}_4 + V^{\prime A} \gamma_A + G^{\prime AB} s_{AB},
\end{split}
\end{equation}
where the transformed quantities are:
\begin{equation} a^\prime = -a, \quad V^\prime = \mathcal{P}_V V, \quad G^\prime = -\mathcal{P}_V G\mathcal{P}^{\intercal}_V.
\end{equation}
with $\mathcal{P}_V$ the vector representation of the parity operator $\mathcal{P}_V = \operatorname{diag}\left(+1, -1, -1, -1, +1 \right)$.
Doing the same for time reversal gives an overall decomposition of\>\footnote{\label{laternote}In cases where $a$, $V$, $G$ are complex, the imaginary parts will have additional oddness under time reversal, but this will not alter the effect of parity. As well, one may swap signature by sending $\gamma_A \mapsto i \gamma_A$. This too can only possibly alter the time reversal properties of the decomposition, leaving charges under parity unchanged.}
\begin{equation}
    \psi \overline{\varphi} = \mathbf{1}_P +\mathbf{5}_T+\mathbf{10}_P.
\end{equation}
Here the subscript indicates additional sign changes garnered when compared to the application of either the standard parity or time reversal operators upon a generic tensor of that type. 

In order for terms in our Lagrangian constructed from these bi-linears to be true scalars, the additional charges under the discrete reflection group must be canceled: each term must be coupled to a tensor of the same kind. For the $5$ vector, we may couple to momentum $\psi \gamma^A P_A \overline{\psi} \equiv \psi i \gamma^A\partial_A \overline{\psi}$, as under time reversal both vectors pick up an additional sign, thus canceling. In order for one to construct a mass term such as $M\overline{\psi}\psi$, the term is not a true scalar unless $M$ is parity odd. While this does invite the idea of a Higgs-type mass mechanism via coupling to a psudeo-scalar field, it foremost demonstrates plainly that theories of $\Spin{1,4}$ spinors cannot have standard Dirac type mass terms.

It is worth mentioning how this relates to the usual Lorentzian theory in four dimensions. In the limit of large de Sitter radius, the de Sitter representations decompose into Lorentzian representations as $\mathbf{5} \mapsto \mathbf{4}\oplus \mathbf{1}$ and $\mathbf{10} \mapsto \mathbf{6} \oplus \mathbf{4}$, which results in an overall Lorentzian spinor bi-linear decomposition of:
\begin{equation}
    \psi \overline{\varphi} = \mathbf{1}_{P}+ \left(\mathbf{4}_{T} +\mathbf{1}_{T}\right) +\left(\mathbf{6}_{P}+\mathbf{4}_{P}\right).
\end{equation}
Because the standard assignment of $\mathcal{P}$ in the Lorentzian theory is different from --- though not incompatible with --- the de Sitter theory ($\gamma^0$ vs $\gamma^0\gamma^5$), these quantities are `shuffled around' compared to the textbook Fierz decomposition. The scalar in the de Sitter case is mapped to the psudeo-scalar in the textbook Lorentzian case, while the mass term in the Lorentzian case can only come from (the imaginary part of) the fifth component of the $5$-vector, painting a rather compelling picture of a theory wherein mass joins the energy-momentum $4$-vector to form a $5$ vector in the $\Spin{1,4}$ picture.
\section{Summary}
A co-representation of the de Sitter pin group $\Pin{4,1}$ was built in order to investigate the parity and time reversal properties of fermion bilinears made from spinors respecting the symmetry group $\Spin{1,4}$. It was shown in this framework the standard Dirac mass terms of the type $m \overline{\psi}\psi$ enter the lagrangian as psudeo-scalar terms under parity. It is noted that despite this, a theory of massless fermions obeying de Sitter symmetry, in the limit of large de Sitter radius may appear locally as a Lorentzian theory of massive fermions. 

%\appendix
%\acknowledgments

% Bibliography
\bibliographystyle{JHEP}
\bibliography{McRae_dS_DiracMass_biblio.bib}

\end{document}